\documentclass[journal=jacsat,manuscript=article]{achemso}
\usepackage[version=3]{mhchem}
\usepackage{chemformula}
\usepackage{array,multirow}
\usepackage{graphics}
\usepackage{comment}

\renewcommand{\thetable}{S\arabic{table}}

\usepackage{caption}
\usepackage{subcaption}
\newcolumntype{P}[1]{>{\centering\arraybackslash}p{#1}}
\usepackage{bm}

\author{Guilherme B. Kanegae} \affiliation{Universidade Estadual de
  Campinas, Instituto de F\'{i}sica Gleb Wataghin, Departamento de
  F\'{i}sica Aplicada, Campinas, S\~ao Paulo, Brazil.}

\author{Marcelo L. Pereira Junior}
\affiliation{University of Bras\'{i}lia, Faculty of Technology, Department of Electrical Engineering, Bras\'{i}lia, Brazil.}

\author{Douglas S. Galv\~{a}o}
\affiliation{Universidade Estadual de Campinas, Instituto de
F\'{i}sica Gleb Wataghin, Departamento de F\'{i}sica Aplicada, Campinas, S\~ao Paulo, Brazil.}

\author{Luiz A. Ribeiro Junior}
\affiliation{University of Bras\'{i}lia, Institute of Physics, Bras\'{i}lia, Brazil.}
\alsoaffiliation{Computational Materials Laboratory, LCCMat, University of Bras\'{i}lia, Bras\'{i}lia, Brazil.}

\author{Alexandre F. Fonseca}
\affiliation{Universidade Estadual de Campinas, Instituto de
F\'{i}sica Gleb Wataghin, Departamento de F\'{i}sica Aplicada, Campinas, S\~ao Paulo, Brazil.}
\alsoaffiliation{Department of Mechanical and Civil Engineering, California Institute of Technology, 1200 E. California Blvd., Pasadena, CA, 91125, USA.}
\email{afonseca@ifi.unicamp.br}

\title
[Enhanced Elastocaloric Effects in $\mathbf\gamma$-graphyne]
{Enhanced Elastocaloric Effects in $\mathbf{\gamma}$-graphyne}


\begin{document}

\begin{tocentry}

\includegraphics[width=\linewidth]{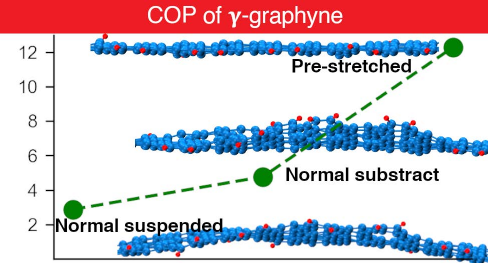}

\end{tocentry}

\begin{abstract}
The global emphasis on sustainable technologies has become a paramount concern for nations worldwide. Specifically, numerous sustainable methods are being explored as promising alternatives to the well-established vapor-compression technologies in cooling and heating devices. One such avenue gaining traction within the scientific community is the elastocaloric effect (eC). This phenomenon holds promise for efficient cooling and heating processes without causing environmental harm. Studies carried out at the nanoscale have demonstrated the efficiency of the eC, proving to be comparable to that of state-of-the-art macroscopic systems. In this study, we used classical molecular dynamics simulations to investigate the elastocaloric effect for $\gamma$-graphyne. Our analysis goes beyond obtaining changes in eC temperature and the coefficient of performance (COP) for two species of $\gamma$-graphyne nanoribbons (armchair and zigzag). We also explore their dependence on various conditions, including whether they are on deposited on a substrate or pre-strained. Our findings reveal a substantial enhancement in the elastocaloric effect for $\gamma$-graphyne nanoribbons when subjected to pre-strain, amplifying it by at least one order of magnitude. Under certain conditions, the change in the eC temperature and the COP of the structures reach expressive values as high as 224 K and 14, respectively. We discuss the implications of these results by examining the shape and behavior of the carbon-carbon bond lengths within the structures.
\end{abstract}

\section{Introduction}

Growing concerns about climate change and the push for sustainable development of nations are steering scientific and technological research \cite{Glass2019}. Addressing the escalating global demand for cooling while mitigating greenhouse gas emissions presents a significant challenge for materials scientists and engineers \cite{dong2021review}. Understanding material properties is crucial for technological progress toward these efforts. It may provide the most direct route to meeting these diverse demands.

In recent decades, the unique properties of nanomaterials (such as graphene \cite{novoselov2004electric}) have been widely exploited, playing a fundamental role in advancing several new technologies across various domains \cite{pandey2022role}. Despite concerns about potential environmental hazards associated with nanomaterials (such as nanoparticles)  \cite{el2023nanomaterials,moon2024SciAdv}, the literature emphasizes the substantial benefits of nanotechnology in critical areas crucial for sustainable human development \cite{Pokrajac2021ACSNano,hussein2023state,saleh2023synthesis}.
 
A key concern in sustainability is the development of environmentally friendly cooling devices \cite{debnath2014environmental}. Material scientists are exploring alternatives to the conventional vapor-compression cooling technology, leading to the emergence of solid-state refrigeration \cite{manosa2013,moya2014}. Based on the caloric effect, this approach involves generating or absorbing heat in response to an external field, which can be electrical, magnetic, or mechanical \cite{chauhan2015review,cazorla2019novel}. In the mechanical category, barocaloric (bC) \cite{li2020JCMA,muniz2020,li2023predicting,Carlos2024npj} and elastocaloric (eC) \cite{xie2017PLA,ray2019science}  effects are prominent, allowing a system to change its temperature through hydrostatic and stress deformations, respectively. Importantly, efficient elastocaloric materials include natural rubber~\cite{muniz2020,ray2019science}, Ni-Ti ~\cite{cui2012apl,cong2019PRL} and Cu-Zn-Al shape memory alloys~\cite{manosa2013apl}, among others~\cite{qian2016review,frenzel2018MRSBull}.

Developing solid-state refrigeration offers distinct advantages, such as potential systems or application miniaturization \cite{bruederlin2018elastocaloric,imran2021}. While computational studies on the eC effect in nanomaterials are still in their early stages, noteworthy findings have already been obtained \cite{lisenkov2016elastocaloric,zhang2017elastocaloric,patel2021elastocaloric,cantuario2019high,silva2022high,ribeiro2023elastocaloric,zhao2022room,cai2023origami}. Lisenkov et al.\cite{lisenkov2016elastocaloric} pioneered molecular dynamics-based predictions for eC effects in graphene and carbon nanotubes. Similarly, Zhang \cite{zhang2017elastocaloric} explored eC effects in boron-nitride nanotubes, and Patel and Kumar \cite{patel2021elastocaloric} predicted the eC effect in zinc-oxide nanowires.

Regarding the methodological aspects of the eC effect, Cantuario and Fonseca \cite{cantuario2019high} introduced a novel protocol to simulate complete thermodynamic cycles of refrigeration. Their work estimated the coefficient of performance (COP) of carbon nanotubes as solid nano refrigerants, demonstrating comparable or superior efficiency to current vapor-compression refrigerators. In addition, Silva and Fonseca~\cite{silva2022high} extended these calculations to determine the COP of different carbon nanotubes across a broad range of operating temperatures. A similar method was applied to unveil eC behaviors in graphene kirigami \cite{ribeiro2023elastocaloric}. Utilizing different protocols, Zhao, Guo, and Zhang~\cite{zhao2022room} and Cai, Yang, and Akbarzadeh~\cite{cai2023origami} predicted significant COP in 3D graphene architectures and graphene origami, respectively. However, to our knowledge, the eC effect in graphyne-based structures remains to be fully explored.

In 1987, Baughman, Eckhardt, and Kertesz introduced porous two-dimensional carbon phases known as {\it graphynes}, proposing structures formed by replacing carbon-carbon bonds in a graphene hexagonal network with acetylene chains \ch{\bond{single}C+C\bond{single}}$_n$\cite{baughman1987structure}. For $n>1$, these structures are termed graph{\it di}yne ($n=2$), graph{\it tri}yne ($n=3$), etc. Like graphene, graphynes are one-atom thick, predicted to exhibit high carrier mobility \cite{chem2013JPCLett} and robust mechanical strength \cite{cranford2011Carbon}. Distinguishingly, graphynes are porous, having a direct electronic bandgap \cite{malko2012PRL} and tunable electronic behavior \cite{padilha2014JPCC}.

Graphynes exhibit unique properties, including negative thermal expansion \cite{Hernandez2017DRM,mondal2023PCCP}, high Poisson's ratio \cite{Hernandez2017DRM,kanegae2022effective}, negative linear compressibility \cite{kanegae2022effective}, and thermoelectric capabilities \cite{Sevik2014apl}. Mechanical properties have been explored concerning $n$ or density variations \cite{cranford2012nanoscale,hou2015JAM,kanegae2022effective,kanegae2023density}. Computational investigations have extended to graphyne scrolls \cite{douglas2018graphynescrool}, nanoribbons~\cite{Liu2024PCCP}, and nanotubes \cite{coluci2003families,coluci2004theoretical,de2019elastic}.

After multiple attempts to produce acetylenic oligomers \cite{die1994nature,kehoe2000carbon,haley2008PAC,die2010AdvM}, Li \textit{et al.}\cite{li2010architecture} achieved a breakthrough by synthesizing the most symmetric form of graphyne with $n=2$, known as $\gamma$-graphdiyne. Subsequently, two groups reported the synthesis of $n=4$ graphtetraynes eight to ten years later\cite{gao2018nanoenergy,pan2020chem}. Remarkably, in 2022, three independent groups reported scalable syntheses of the most symmetric and densest form of $n=1$ graphyne, the $\gamma$-graphyne \cite{hu2022synthesis,desyatkin2022scalable,barua2022carbon}. These synthetic achievements are of significant importance. The possibility of large-quantity graphyne synthesis poses graphynes as a potential candidate to replace graphene in some applications \cite{site}. $\gamma$-Graphyne has already found diverse applications, such as desalinators \cite{mehrdad2019efficient,azamat2020atomistic}, photocatalyst \cite{lin2020gama}, supercapacitors \cite{chen2020first}, environmental remediation~\cite{majidi2021efficient,zhan2022hollow}, and sensors \cite{nikmanesh2021novel}, among others \cite{narang2023review,li2023Giant}. The synthesis of the densest form of graphyne opens up new possibilities for innovative applications.

In this work, we have used classical reactive molecular dynamics (MD) simulations to investigate the eC effects in monolayers of $\gamma$-graphyne. Using a previously defined Otto-like thermodynamic cycle \cite{cantuario2019high}, our computational protocol provides adiabatic temperature changes and COP values for the cycle's cooling and heating phases. This investigation explores the role of the substrate, boundary conditions, and a specific regime of initial pre-elongation on the intensity and efficiency of the eC in $\gamma$-graphyne. By offering the first estimate for the COP of graphyne as solid refrigerants for both cooling and heating applications, this research fills a gap in the literature concerning this material.

\section{Methodology}

Fully atomistic MD simulations, using the adaptive intermolecular reactive empirical bond order (AIREBO) potential~\cite{brenner,stuart_JCP} were carried out to investigate the effects of eC on $\gamma$-graphyne nanoribbons. The AIREBO potential is well-known for its effectiveness in reliably describing different carbon nanostructures \cite{cho2002NL,shenoy2010Science,peeters2011PRB,muniz2015JCCP,muniz2018ACSAMI}. The simulations were designed to determine the eC temperature changes and COP magnitudes. It should be noted that AIREBO has already been widely adopted for studying the structure, mechanical, and thermal properties of graphynes \cite{zhang2012APL,Hernandez2017DRM,Yang2012CMS,Ma2016Carbon,kanegae2022effective}.

\begin{figure}[htb!]
\begin{center}
\includegraphics[width=\linewidth]{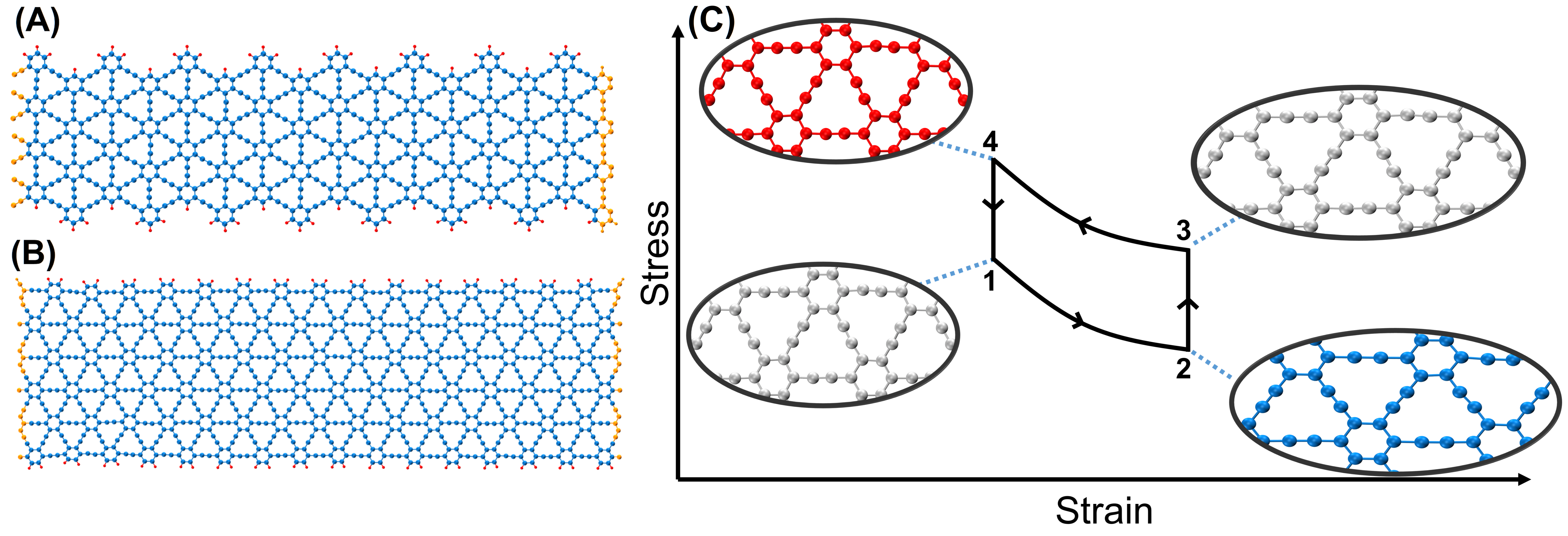}
\caption{This Figure illustrates the zigzag (A) and armchair (B) $\gamma$-graphyne nanoribbons. Free/movable carbon and hydrogen atoms are indicated in blue color. Orange indicates the atoms that are either fixed or rigidly moved in the finite-length simulations. Panel (C) shows the stress-strain thermodynamic cycle, with insets depicting thermodynamic equilibrium states. White, red, and blue colors correspond to temperatures equal to, greater than, and smaller than the operator temperature.}
\label{fig1}
\end{center}
\end{figure}

The simulations were carried out using the Large-Scale Atomic/Molecular Massively Parallel Simulator (LAMMPS) package \cite{plimpton_CPC}. To characterize the eC effect, we performed simulations of the proposed Otto cycle \cite{cantuario2019high}, determining temperature changes ($\Delta T$) and energy quantities necessary for estimating the COP of graphyne nanoribbons. We have investigated the eC of nanoribbons cut along the zigzag and armchair directions. Figure \ref{fig1} illustrates the structures, a qualitative diagram depicting the thermodynamic cycle, and the equilibrium states used to determine the eC in graphynes.

The hydrogen-passivated $\gamma$-graphyne nanoribbons considered in this study cut along the zigzag and armchair directions are depicted in Figs. \ref{fig1}A and \ref{fig1}B, respectively. They exhibit dimensions of approximately $96.5\times29.3$ \AA\mbox{} (zigzag) and $110.0\times34.7$ \AA\mbox{} (armchair). The initial energy minimization was performed for all structures using the LAMMPS package conjugate gradient algorithms to obtain the minimum energy configuration. Energy and force convergence criteria during minimizations were set to $10^{-8}$ eV/\AA\mbox{} and $10^{-8}$ eV/\AA\mbox{}, respectively, with convergence ensured following the protocols by Sihn {\it et al.} \cite{Sihn2015Carbon}.

For the infinite structures (periodic boundary conditions), equilibrium at zero strain and initial operating temperature ($T_0=300$ K) was achieved through MD simulation lasting 1 ns. A timestep of $0.5$ fs and a fixed number of atoms ($N$), temperature ($T_0$), and pressure ($P=0$) were maintained. Atom positions and the nanoribbon box size along the length direction were allowed to equilibrate.

For the finite structures, equilibrium at zero strain and initial operating temperature ($T_0=300$ K) was attained through a 60 ps of MD simulation. A timestep of $0.5$ fs, a Langevin thermostat at $T_0$, and the constraint of allowing only the atoms at the extremities (orange atoms in Fig. \ref{fig1}) to move along the length direction of the nanoribbon were applied to equilibrate the positions of the atoms and the length of the structure.

These simulations were performed with the LAMMPS parameters PDAMP = 0.5 ps and TDAMP = 1 ps. Then, the equilibrium length, $L_0$, is taken from the average structure length values during the last 20 ps of the simulations. $L_0$ defines the zero strain state of the structures. 

The thermodynamic cycle, shown in Fig. \ref{fig1}C, has four paths through equilibrium states numbered 1 to 4. State number 1 represents the structure at the initial deformation configuration (which may not always be zero, as explained ahead) and the operating temperature, $T_0$. Two paths are isostrain (${4 \rightarrow 1}$ and ${2 \rightarrow 3}$), and two are adiabatic (${1 \rightarrow 2}$ and ${3 \rightarrow 4}$). Isostrain paths involve attaching a thermal bath to the system at the operating temperature, $T_0=300$ K. The two adiabatic routes increase and decrease the tensile strain between $\varepsilon_i$ and $\varepsilon_i + 0.1$.

The simulation protocols for the thermodynamic cycles are as follows: For isostrain paths, we perform simulations on a total of 60 ps with a time step of $0.5$ fs. We maintain the strain (or fixed structure size) and apply a Langevin thermostat with TDAMP = 1 ps. For adiabatic tensile and tensile release strain paths, we start the simulations with a $T_0=300$ K Boltzmann distribution of velocities to ensure that the initial temperature aligns with the operating one. Subsequently, we apply a strain rate of 0.1 \%ps$^{-1}$ for 100 ps. These simulations proceed without a thermostat and with a timestep of, at least, $0.05$ fs. We set the final strain to be 10\% above the initial strain. Recognizing the potential impact of the strain rate on the eC results, we present preliminary tests in Section {\bf S1} of the Supporting Information (SI) to assess the convergence of the COP results for different strain rates, providing further information.

The following combinations of conditions were considered for each $\gamma$-graphyne nanoribbon in this study: 

\begin{enumerate}
    
\item[(i)] two pre-elongation regimes, i.e., two thermodynamic cycles with one starting from zero strain ($\varepsilon_i = 0$) and the other starting at a specific value ($\varepsilon_i = \varepsilon_0$); \label{i}

\item[(ii)] two boundary conditions, i.e., the nanoribbons will be simulated with periodic boundary conditions (PBC) and finite length (FL); \label{ii}

\item[(iii)] two environmental conditions, i.e., suspended nanoribbons in space (SUS) and nanoribbons deposited on a substrate (SUBS). \label{iii}

\end{enumerate}

Combinations of these conditions result in 16 groups of simulations ($2 \times 2 \times 2 = 16$), hereafter referred to as ``groups''. Each group involves performing 10 simulations that satisfy the same (i), (ii), and (iii) conditions. However, each simulation starts from a statistically (microcanonically) different initial condition corresponding to the same macroscopic condition. This diversity is achieved by providing distinct initial velocity distributions among the system's atoms while maintaining the same $T_0$.

To determine the values of $\varepsilon_0$ for simulations under the second condition (i), we use the eC temperature changing profile during the expansion path of the thermodynamic cycles with $\varepsilon_i = 0$. Within each group, we obtain 10 different values of $\varepsilon_{0j}$ and calculate the corresponding $\varepsilon_0$ for the group as the average over these 10 values. We derive each $\varepsilon_{0j}$ by fitting a previously smooth-averaged $T$ versus $\varepsilon$ curve, corresponding to the expansion path of a single simulation within the group, to a second-degree polynomial with four parameters: $y = a(x+b)^2 + cx + d$. Section {\bf S2.1} in SI provides an illustrative example of determining one $\varepsilon_{0j}$. In contrast, Section {\bf S3} of SI displays typical examples of 10 microcanonically different but macroscopically equivalent temperature profiles during the adiabatic paths.

The substrate was simulated through an infinite artificial wall that interacts with the nanoribbons through a 12-6 Lennard-Jones function with $\epsilon = 0.002844$ eV, $\sigma = 2.9$ \AA, and a cutoff of 10.0 \AA. These parameters lead the equilibrium substrate-nanoribbons distance to be 3.4 \AA. 

As the classical MD simulations do not explicitly capture the quantum effects on the heat capacity of the structures, we use the following relationship to obtain the real eC temperature change, $\Delta T_{\mbox{\scriptsize{REAL}}}$:

\begin{equation}
  \label{eq1}
     Q = C_{\mbox{\scriptsize{MD}}}\Delta T_{\mbox{\scriptsize{MD}}} = C_{\mbox{\scriptsize{REAL}}}\Delta T_{\mbox{\scriptsize{REAL}}}.
\end{equation}

\noindent The Equation (\ref{eq1}) is expressed as $C_{\mbox{\scriptsize{MD}}}$ and $\Delta T_{\mbox{\scriptsize{MD}}}$ representing the heat capacity and eC a temperature change of the structure obtained from MD simulations, respectively. Additionally, $C_{\mbox{\scriptsize{REAL}}}$ signifies the real heat capacity of graphyne at the operating temperature, $T_0$. In the second equality, both sides of Equation (\ref{eq1}) can be divided by the sample mass, making it applicable for $C$ to represent the specific heat in J/(gK) instead of heat capacity. The value of $C_{\mbox{\scriptsize{REAL}}}$ for graphyne was obtained from density functional theory (DFT) calculations carried out as follows.

Calculating constant-volume heat capacities for $\gamma$-graphyne nanoribbons involved using the Quantum Espresso (QE) code \cite{giannozzi2009quantum,giannozzi2017advanced}. The QE-DFT calculations used a plane-wave basis set and pseudopotentials method. The interactions between electrons and atomic cores are described by expanding the wave function using plane-wave and norm-conserving pseudopotentials. Norm-conserving pseudopotentials, generated using the Troullier-Martins method \cite{troullier1991efficient} with the Perdew-Burke-Ernzerhof (PBE) functional, were used to account for both exchange energies and electronic exchange-correlation effects \cite{perdew1996generalized,ernzerhof1999assessment}.

We applied the Broyden-Fletcher-Goldfarb-Shanno (BFGS) scheme to optimize the crystal structure in our model sheets, as implemented in the QE code \cite{head1985broyden}. We adopted high cutoff values, set at 70 Ry for wave functions and 700 Ry for charge density. Electronic self-consistency was achieved with a convergence criterion of 1.0 $\times$ 10$^{-5}$ eV. Throughout lattice relaxation, the force on each atom was kept below 1.0 $\times$ 10$^{-3}$ eV/\r{A}.

We used the Phonopy package \cite{togo2015first} alongside QE to address the lattice thermal properties of the studied materials. Phonons, serving as manifestations of vibrational modes within a crystal lattice, play a pivotal role in determining the thermal characteristics of materials. The seamless integration of Phonopy with QE allows precise computing of the force constants and provides essential insights for understanding phonon analyses and lattice thermal properties. For the computation of second and third-order interatomic force constants, we applied the supercell finite-displacement approach with step sizes of 1.0$\times10^{-2}$ \r{A}. Supercells containing 24 atoms were explicitly employed to calculate the second-order interatomic force constants.

The COP corresponding to the cooling or heating device is given by:

\begin{equation}
  \label{eqCOP}
     \mbox{COP} = \frac{Q_{\mbox{\scriptsize{C}}}\mbox{ or } Q_{\mbox{\scriptsize{H}}}}{W}.
\end{equation}

In this expression, $Q_{\mbox{\scriptsize{C}}}$ (or $Q_{\mbox{\scriptsize{H}}}$) and $W$ denote the heat exchanged with the thermal bath after the cooling/expansion (or heating/contraction) paths and the total work per thermodynamic cycle, respectively. We can obtain $Q_{\mbox{\scriptsize{C}}}$ or $Q_{\mbox{\scriptsize{H}}}$ by multiplying the heat capacity by the corresponding $\Delta T$ since the work done by or on the system is null during the isostrain paths. No heat is exchanged during the adiabatic paths; then the total work, $W$, done on the system during one thermodynamic cycle can be computed by summing the total internal energy differences of the system between thermodynamic states 2 and 1 ($W_{\mbox{\scriptsize{EXPANSION}}}=E_2 - E_1$) and states 4 and 3 ($W_{\mbox{\scriptsize{CONTRACTION}}}=E_4 - E_3$), i.e., 

\begin{equation}
  \label{eqW}
     W = W_{\mbox{\scriptsize{EXPANSION}}} + W_{\mbox{\scriptsize{CONTRACTION}}}\, .
\end{equation}

\section{Results}

\begin{figure}[htb!]
\begin{center}
\includegraphics[width=\linewidth]{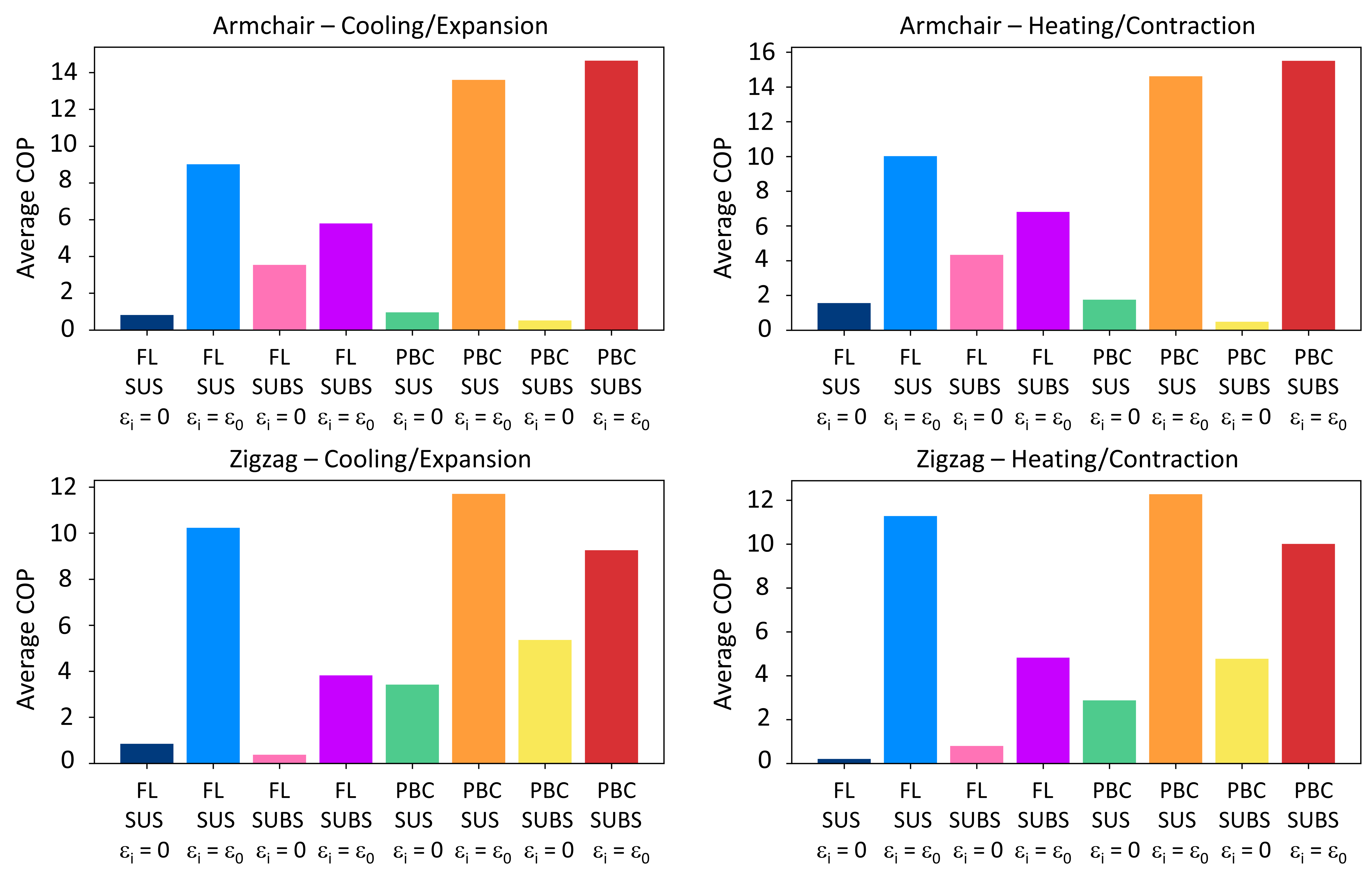}
\caption{This Figure presents the average COP values across different simulated conditions for all structures during cooling/expansion and heating/contraction. The labels FL, PBC, SUS, and SUBS correspond to ``finite length'', ``periodic boundary conditions'', ``suspended'', and ``substrate'', respectively. For the definitions of these conditions, refer to the previous section. $\varepsilon_i$ represents the strain of the initial state in the thermodynamic cycle.}
\label{fig2}
\end{center}
\end{figure}

In Table \ref{tab-T} and Figure \ref{fig2}, we present the averaged real eC temperature changes, $\Delta T_{\mbox{\scriptsize{REAL}}}$, and the averaged COP for all structures simulated under different conditions. Determining $\Delta T_{\mbox{\scriptsize{REAL}}}$ using equation (\ref{eq1}) involves calculating the real and MD-specific heats of all structures. The real specific heat of $\gamma$-graphyne is considered as the one obtained from DFT calculations, denoted by $c_{\mbox{\tiny{REAL}}} = 0.4383$ J/gK. The MD values of specific heat are detailed in Table {\bf S4} of the Supporting Information (SI). Additionally, Section {\bf S4} of SI provides two examples of a series of temperature equilibrations of the structures, ranging from 50 to 700 K, and their corresponding temperature and energy profiles necessary for determining the specific heat.

\begin{table}[htb]
  \caption{The real eC temperature changes ($\Delta T_{\mbox{\scriptsize{REAL}}}$ in K), averaged over the 10 statistically different simulations within each group of conditions, for the armchair (AC) and zigzag (ZZ) structures during the expansion and contraction paths of the thermodynamic cycle. The terms ``Finite Length'', ``Periodic Boundary Conditions'', ``Suspended'', and ``Substrate'' are defined in the Methodology section.}
  \label{tab-T}
  \resizebox{\columnwidth}{!}{%
\begin{tabular}{clcclcclcclccl}
\hline
\hline
\hline
\multirow{2}{*}{Conditions} & & \multicolumn{11}{c}{Finite Length}                                                                                                                                                                                                &  \\\cline{3-13}
                            & & \multicolumn{5}{c}{Suspended}                                                                              & & \multicolumn{5}{c}{Substrate}                                                                             &  \\
                            \cline{1-1}  \cline{3-7} \cline{9-13} 
Chirality                   & & \multicolumn{2}{c}{AC}                                & & \multicolumn{2}{c}{ZZ}                                & & \multicolumn{2}{c}{AC}                                & & \multicolumn{2}{c}{ZZ}                                &  \\ \cline{1-1} \cline{3-4}  \cline{6-7}  \cline{9-10}  \cline{12-13}  
Initial Strain              & & $\varepsilon_i = 0$ & $\varepsilon_i = \varepsilon_0$ & & $\varepsilon_i = 0$ & $\varepsilon_i = \varepsilon_0$ & & $\varepsilon_i = 0$ & $\varepsilon_i = \varepsilon_0$ & & $\varepsilon_i = 0$ & $\varepsilon_i = \varepsilon_0$ &  \\ \hline
Expansion                    & & 15.5                & 124.6                           & & 23.8                & 184.1                           & & 39.9                & 118.1                           & & 8.3                 & 170.4                           &  \\
Contraction                 & & 34.2                & 139.2                           & & 4.8                 & 203.2                           & & 49.9                & 140.7                           & & 15.8                & 224.6                           &  \\ \hline
\hline
\hline
\multirow{2}{*}{Conditions} & & \multicolumn{11}{c}{Periodic Boundary Conditions}                                                                                                                                                                                           &  \\\cline{3-13}
                            & & \multicolumn{5}{c}{Suspended}                                                                              & & \multicolumn{5}{c}{Substrate}                                                                             &  \\ \cline{1-1}  \cline{3-7} \cline{9-13} 
Chirality                   & & \multicolumn{2}{c}{AC}                                & & \multicolumn{2}{c}{ZZ}                                & & \multicolumn{2}{c}{AC}                                & & \multicolumn{2}{c}{ZZ}                                &  \\ \cline{1-1} \cline{3-4}  \cline{6-7}  \cline{9-10}  \cline{12-13}  
Initial Strain              & & $\varepsilon_i = 0$ & $\varepsilon_i = \varepsilon_0$ & & $\varepsilon_i = 0$ & $\varepsilon_i = \varepsilon_0$ & & $\varepsilon_i = 0$ & $\varepsilon_i = \varepsilon_0$ & & $\varepsilon_i = 0$ & $\varepsilon_i = \varepsilon_0$ &  \\ \hline
Expansion                    & & 11.3                & 81.2                            & & 42.9                & 154.6                           & & 1.72                & 83.3                            & & 60.3                & 153.8                           &  \\
Contraction                 & & 21.0                & 87.3                            & & 33.9                & 162.8                           & & 1.91                & 88.1                            & & 52.8                & 166.0                           & \\ \hline
    \hline
    \hline
\end{tabular}
}
\end{table}

From Table \ref{tab-T}, we can see that pre-strained structures exhibit eC temperature changes at least one order of magnitude larger than initially unstrained structures, both during expansion and contraction. This trend is consistent with observations for the eC of pre-strained rubber \cite{Guyomar2015APL}, where an approximately 50\% increase in the eC temperature change was observed. The most significant eC temperature changes in this study are observed for the finite-length suspended zigzag structure after the expansion path ($\Delta T_{\mbox{\scriptsize{REAL}}}=184.1$ K) and for the finite-length zigzag structure on the substrate after the contraction path ($\Delta T_{\mbox{\scriptsize{REAL}}}=224.6$ K).

Except for armchair structures under PBC (for both expansion and contraction paths) and finite-length zigzag structures (for expansion paths only), the enhancements in the eC temperature change by simply placing the structure on a substrate for unstrained ones can reach up to three times. Similar improvements in eC temperature change due to substrate adhesion were observed by Meng {\it et al.}~\cite{li2020adhesion} for graphene, of the order of 30\%. For pre-strained structures, there is no significant difference in the eC temperature change between suspended and deposited on substrate $\gamma$-graphyne nanoribbons.

The temperature profiles presented in Figures from {\bf S3.1} to {\bf S3.4} reveal that during expansion (contraction), the eC temperature change corresponds to cooling (heating). This behavior is consistent with the observations made by Lisenkov {\it et al.}\cite{lisenkov2016elastocaloric}, Cantuario and Fonseca\cite{cantuario2019high}, Meng {\it et al.}\cite{li2020adhesion}, and Silva and Fonseca\cite{silva2022high} for the eC of carbon nanotubes and graphene. However, this behavior is the opposite of the observed in rubber~\cite{Guyomar2015APL} and kirigami graphene~\cite{ribeiro2023elastocaloric}, where the eC temperature change is positive when stretched and negative when the strain is released.

The COP is as essential as the amount of eC temperature change, as it allows for comparing the efficiency of different refrigerators and heat pumps. While the best COP values of vapor-compression-based refrigerators are around 2.2~\cite{nada2023IJT}, determining the COP for the structures at each set of conditions (or group) involved averaging the COP from 10 statistically different simulations under the same macroscopic conditions. Instead of presenting the COP in a table, representing it in column bars is more effective for direct comparisons. In Figure \ref{fig2}, we show the COP values for all structures under all conditions considered in the present study. Like the eC temperature changes, pre-strained structures exhibit much larger COP values than initially relaxed ones.

Figure \ref{fig2} also indicates that, for unstressed armchair finite length and zigzag PBC structures, depositing the $\gamma$-graphyne nanoribbon on a substrate enhances the elastocaloric efficiency compared to the suspended case, in agreement with the findings of Meng {\it et al.}~\cite{li2020adhesion}. 

\begin{figure}[htb!]
\begin{center}
\includegraphics[width=\linewidth]{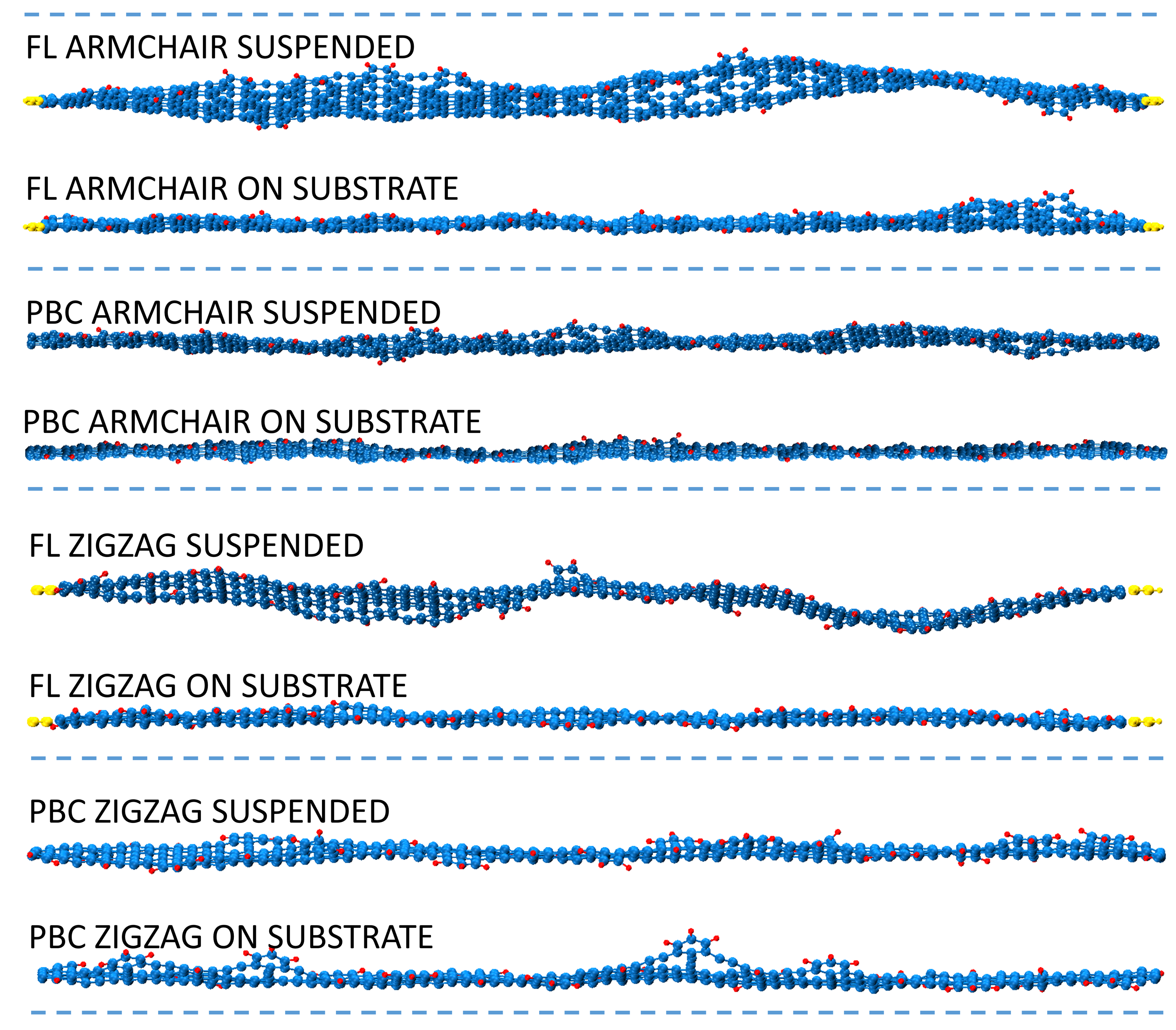}
\caption{Lateral views of the armchair and zigzag structures on different boundary (ii) and environment (iii) conditions. FL and PBC denote ``finite length'' and ``periodic boundary conditions''. Refer to the previous section for the meanings of these conditions. Dashed lines separate structures for each pair of environmental (iii) conditions. Movable carbon and hydrogen atoms, and fixed carbon ones are shown in blue, red, and yellow colors, respectively.}
\label{fig3}
\end{center}
\end{figure}

To gain further insights into the significant increase in the efficiency of the elastocaloric-based refrigerator/heater with pre-straining, we examined in detail the structures in their relaxed state. Lateral views of the equilibrated structures without pre-strain are shown in Figure \ref{fig3} for all boundary (ii) and environment (iii) conditions.

Figure \ref{fig3} illustrates that unstressed suspended structures exhibit a much more wavy-like lateral form than those on the substrate. When a tensile strain is applied to the nanoribbon, its undulations become initially aligned. Then, its bonds start to be strained. We conjecture that this might be related to the increase and subsequent decrease of the eC temperature profile during both expansion and compression in unstrained structures. See the examples in Figures {\bf S3.1} and {\bf S3.3}. 

The straining of initially unstrained $\gamma$-graphyne nanoribbons during the expansion path can be divided into two phases. The first phase involves two effects: alignment of the nanoribbon chains with the direction of strain and some increase in carbon-carbon bond length values. The second phase comprises only bond length increase but at a slightly more significant rate. Only the second phase occurs when beginning the thermodynamic cycle from a pre-strained structure. For this case, the eC temperature change exhibits a monotonically decreasing (increasing) behavior when subjected to tensile strain (tensile strain release), as shown in the examples given in Figures {\bf S3.2} and {\bf S3.4}. 

\begin{figure}[htb!]
\begin{center}
\includegraphics[width=\linewidth]{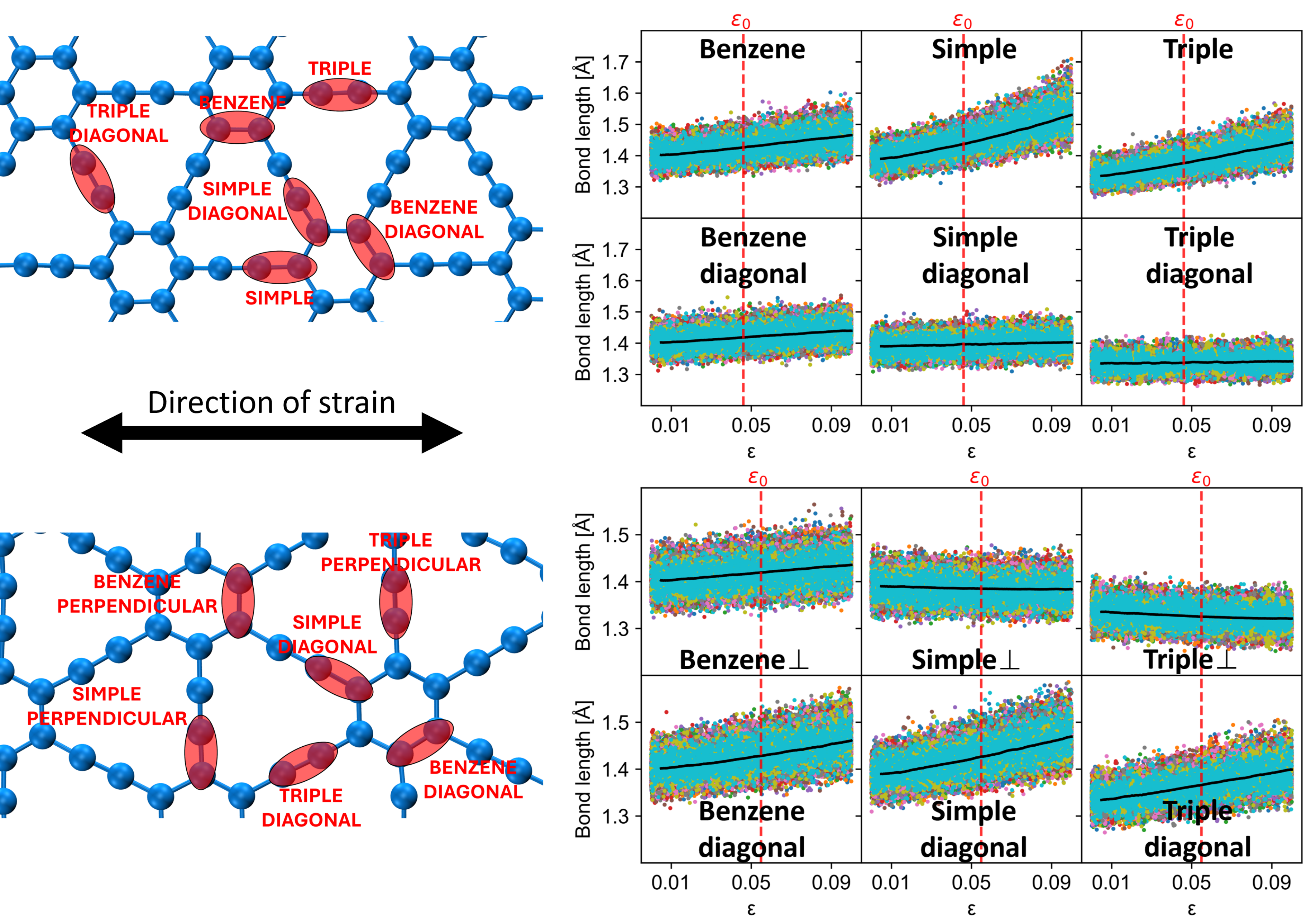}
\caption{This Figure illustrates the evolution of the bond length values of six different carbon-carbon bonds in unstrained armchair (top) and zigzag (bottom) suspended structures. The highlighted sections of the armchair and zigzag structures on the left indicate these six bonds. Each type of bond length was measured in 30 different regions along the structure, plotting the strain values at intervals of $5\times10^{-5}$. Each dot color corresponds to one of these regions. The black curves represent the average smooth-average curves corresponding to the bond length values in one region. The symbol $\bot$ denotes ``perpendicular''. Vertical dashed lines mark the corresponding value of $\varepsilon_0$. The green shades show the range of fluctuations in each measure.}
\label{fig4}
\end{center}
\end{figure}

Finally, to offer quantitative support for the above conjecture, we analyzed the evolution of the length of various carbon-carbon bonds in unstrained armchair and zigzag suspended structures during the expansion path. The results are presented in Figure \ref{fig4}.

The carbon-carbon bonds primarily bearing tension in the armchair structure are aligned directly with the strain direction. The diagonal carbon-carbon bonds are the ones that endure most of the tension in the zigzag structure. The most significant bond length increase occurs in the acetylenic simple bond aligned with the strain direction in armchair structures and benzene and acetylenic simple diagonal bonds in zigzag ones.

Upon inspecting the graphs for unstrained armchair structures, it becomes apparent that for $\varepsilon > \varepsilon_0$, the slope of the simple bond length versus $\varepsilon$ curve is slightly higher than that for $\varepsilon < \varepsilon_0$. This difference is not visually discernible in the case of the primary bonds in zigzag structures. To provide more quantitative evidence for this observation, we performed a linear fitting to the average curves of carbon-carbon bond length versus $\varepsilon$ in two regions: $\varepsilon < \varepsilon_0$ and $\varepsilon > \varepsilon_0$. The results are presented in Table \ref{tabinclina}.

\begin{table}
  \caption{Change rates of the bond length values of the different carbon-carbon bonds with $\varepsilon$ (in \AA) for the bonds shown in Figure \ref{fig4}, for armchair (AC) and zigzag (ZZ) structures considered in Figure \ref{fig4}. The results for the bonds that experience the largest strains are highlighted in {\bf bold}. The types of the bonds are shown in Figure \ref{fig4}. The symbol $\bot$ means ``perpendicular''.}
  \label{tabinclina}
  \begin{tabular}{ccc}
    \hline
    \hline
    \hline
    Structure and type of bond  &  Bond length rates for $\varepsilon<\varepsilon_0$ & Bond length rates for $\varepsilon>\varepsilon_0$  \\
    \hline
    \bf AC Benzene  & \bf 0.634 & \bf 0.716 \\
    \bf AC Simple   & \bf 1.223 & \bf 1.756 \\
    \bf AC Triple   & \bf 1.033 & \bf 1.225 \\
    AC Benzene diagonal   & 0.421 & 0.415 \\
    AC Simple diagonal   & 0.145 & 0.144 \\
    AC Triple diagonal  & 0.055 & 0.085 \\
    \bf ZZ Benzene diagonal & \bf 0.557 & \bf 0.721 \\
    \bf ZZ Simple diagonal & \bf 0.772 & \bf 0.963 \\
    \bf ZZ Triple diagonal & \bf 0.657 & \bf 0.731 \\
    ZZ Benzene $\bot$  & 0.357 & 0.356 \\
    ZZ Simple $\bot$ & -0.106 & -0.019 \\
    ZZ Triple $\bot$ & -0.204 & -0.087 \\
    \hline
    \hline
    \hline
  \end{tabular}
\end{table}

The results in Table \ref{tabinclina} demonstrate a significant change in the rate of bond length variation with strain at $\varepsilon_0$ for the specific bonds that experience most of the overall structural strain. The difference in the rate is noticeable in the first decimal point for these primary carbon-carbon bonds (highlighted in bold in Table \ref{tabinclina}). Meanwhile, there is virtually no change in the rate of bond length variation for the diagonal bonds in armchair structures and the perpendicular ($\bot$) bonds in zigzag ones. The negative rate of variation of the perpendicular bonds with strain indicates the positive Poisson's ratio nature of the structure. Therefore, Table \ref{tabinclina} results suggest a strong correlation between the rate of carbon-carbon bond length variation and the sign of variation of the eC temperature changes.  

\section{Conclusions}

In summary, we explored the eC properties of nanoribbons of $\gamma$-graphyne along armchair and zigzag directions under various conditions. We simulated thermodynamic cycles, incorporating isostrain and adiabatic paths, for $\gamma$-graphyne structures with different configurations, such as finite length, periodic boundary conditions, suspended, and deposited on a substrate. Our findings indicated that placing the nanoribbon on a substrate for initially unstrained structures enhances its eC temperature change and COP values. However, the most substantial improvements in eC temperature change and COP were observed for initially pre-strained structures, irrespective of whether they were being deposited on a substrate or suspended, finite, or with PBC. Additionally, we carried out a detailed analysis of the evolution of carbon-carbon bond lengths, revealing a correlation between the change in the rate of bond length variation with strain and the sign change in eC temperature changes. 

\begin{acknowledgement}
This work received partial support from Brazilian agencies CAPES, CNPq, FAPESP, and FAPDF. M.L.P.J acknowledges the financial support from FAP-DF grant 00193-00001807/2023-16. L.A.R.J. acknowledges CAPES for partially financing this study-Finance Code 88887.691997/2022-00. CENAPAD-SP (Centro Nacional de Alto Desenpenho em Sao Paulo - Universidade Estadual de Campinas - UNICAMP) provided computational support for L.A.R.J, M.L.P.J, A.F.F., and G.B.K. (proj634, proj960, and proj861). L.A.R.J thanks the financial support from Brazilian Research Council FAP-DF grant 00193.00001808/2022-71, FAPDF-PRONEM grant 00193.00001247/2021-2, 00193-00001857/2023-95, 00193-00001782/2023-42, and CNPq grants 350176/2022-1. G.B.K. acknowledges support from the Coordenacao de Aperfeicoamento de Pessoal de Nivel Superior - Brasil (CAPES) - Finance Code 001. A.F.F. is a fellow of the Brazilian Agency CNPq-Brazil (303284/2021-8) and acknowledges grants \#2020/02044-9 and \#2023/02651-0 from S\~{a}o Paulo Research Foundation (FAPESP). L.A.R.J and M.L.P.J also thank Nucleo de Computacao de Alto Desempenho (NACAD) for computational facilities through the Lobo Carneiro Supercomputer. D. S. Galvao acknowledges support from the FAPESP/CEPID Grant 2013/08293-7.
\end{acknowledgement}


\bibliography{bibliography}

\newpage

\large{ -- Supporting Information --\\Elastocaloric Effects in $\mathbf{\gamma}$-graphyne}


\section{S1. Strain Rate Tests}

Table \ref{tabS1} displays the average coefficient of performance (COP) values, derived from 10 distinct simulation runs of the thermodynamic cycle, for zigzag and armchair $\gamma$-graphyne nanoribbons subjected to various strain rates.\\

\renewcommand{\thetable}{S1.1}
\begin{table}
	\caption{Average values of COPs obtained for different strain rates of adiabatic expansion and contraction paths.}
	\label{tabS1}
	\begin{tabular}{ccccccccc}
		\hline
		Strain Rate & \multicolumn{2}{c}{0.003125} & \multicolumn{2}{c}{0.002000} & \multicolumn{2}{c}{0.001000} & \multicolumn{2}{c}{0.000500} \\ \hline
		 Chirality  &  ZZ  &          AC           &  ZZ  &          AC           &  ZZ  &          AC           &  ZZ  &          AC           \\ \hline
		 Expansion  & 0.27 &         2.06          & 0.30 &         2.61          & 0.37 &         3.54          & 0.48 &         2.62          \\
		Contraction & 0.84 &         2.88          & 1.21 &         3.51          & 0.80 &         4.33          & 0.47 &         3.76          \\ \hline
	\end{tabular}
\end{table}

\section{S2. Initial Strain Determination}

Figure \ref{figS2.1} presents two examples illustrating the determination of $\varepsilon_0$ from the elastocaloric (eC) temperature change profiles during the expansion of armchair and zigzag nanoribbons

\renewcommand{\thefigure}{S2.1}
\begin{figure}
    \centering
    \includegraphics[width=\linewidth]{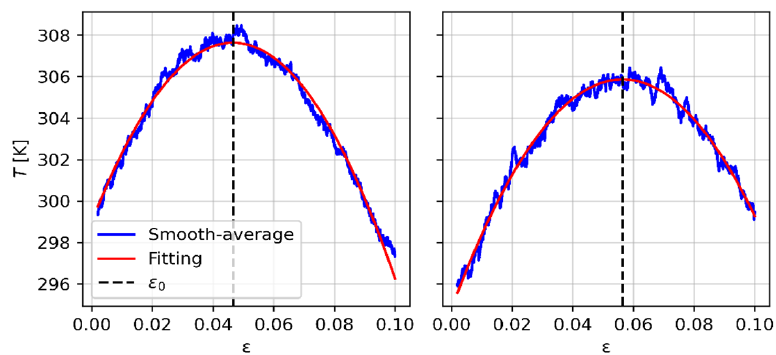}
    \caption{Fitted (red) smooth-averaged (blue) temperature profiles during the expansion of an armchair (left) and zigzag (right) $\gamma$-graphyne. Fitted function: $y = a (x+b)^2 + cx + d$.}
    \label{figS2.1}
\end{figure}

\section{S3. Typical eC Effect Temperature Changing Profiles}

Figures \ref{figS3.1} to \ref{figS3.4} show the eC temperature changing profiles for both expansion (red) and contraction (blue) paths of thermodynamic cycles for armchair $\gamma$-graphyne nanoribbon deposited on a substrate. 

\renewcommand{\thefigure}{S3.1}
\begin{figure}
    \centering
    \includegraphics[width=0.8\linewidth]{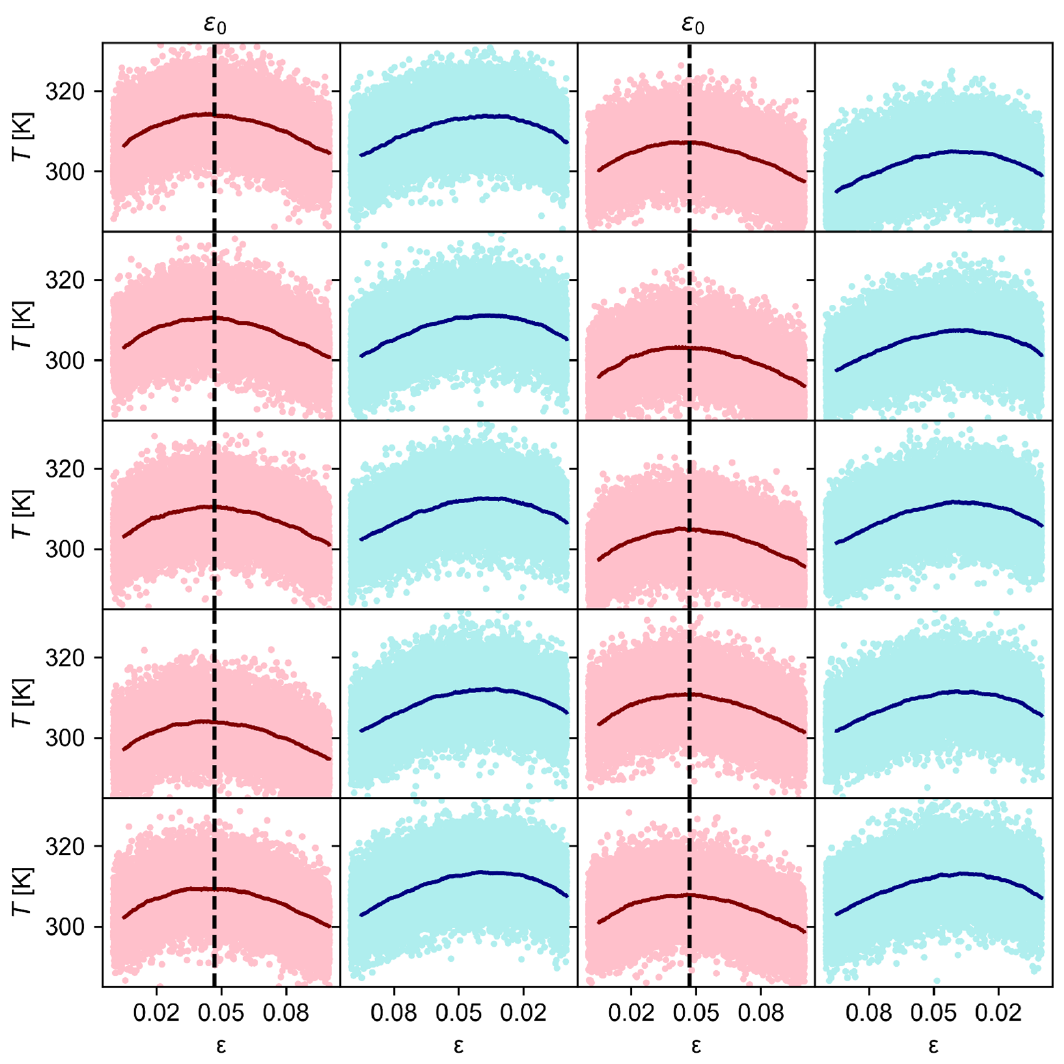}
    \caption{Ten pairs of temperature change profiles of expansion (red) and contraction (blue) paths of the thermodynamic cycles simulated with an armchair $\gamma$-graphyne nanoribbon on a substrate. The strain $\varepsilon_0$ corresponds to the maximum temperature change during expansion.}
    \label{figS3.1}
\end{figure}

\renewcommand{\thefigure}{S3.2}
\begin{figure}
    \centering
    \includegraphics[width=0.8\linewidth]{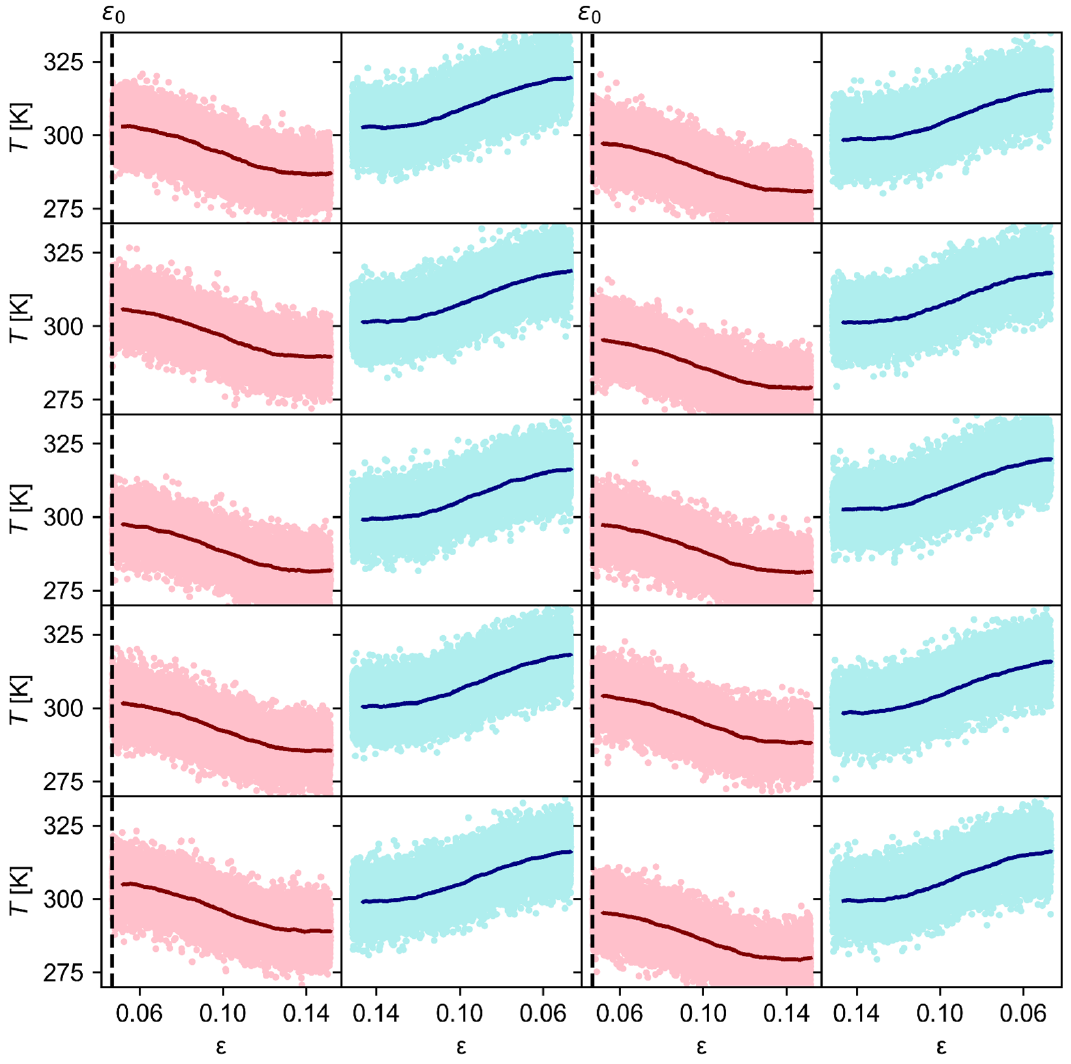}
    \caption{Ten pairs of temperature change profiles of expansion (red) and contraction (blue) paths of the thermodynamic cycles simulated with an armchair $\gamma$-graphyne nanoribbon on a substrate with a pre-elongation of $\varepsilon_0$.}
    \label{figS3.2}
\end{figure}

\renewcommand{\thefigure}{S3.3}
\begin{figure}
    \centering
    \includegraphics[width=0.8\linewidth]{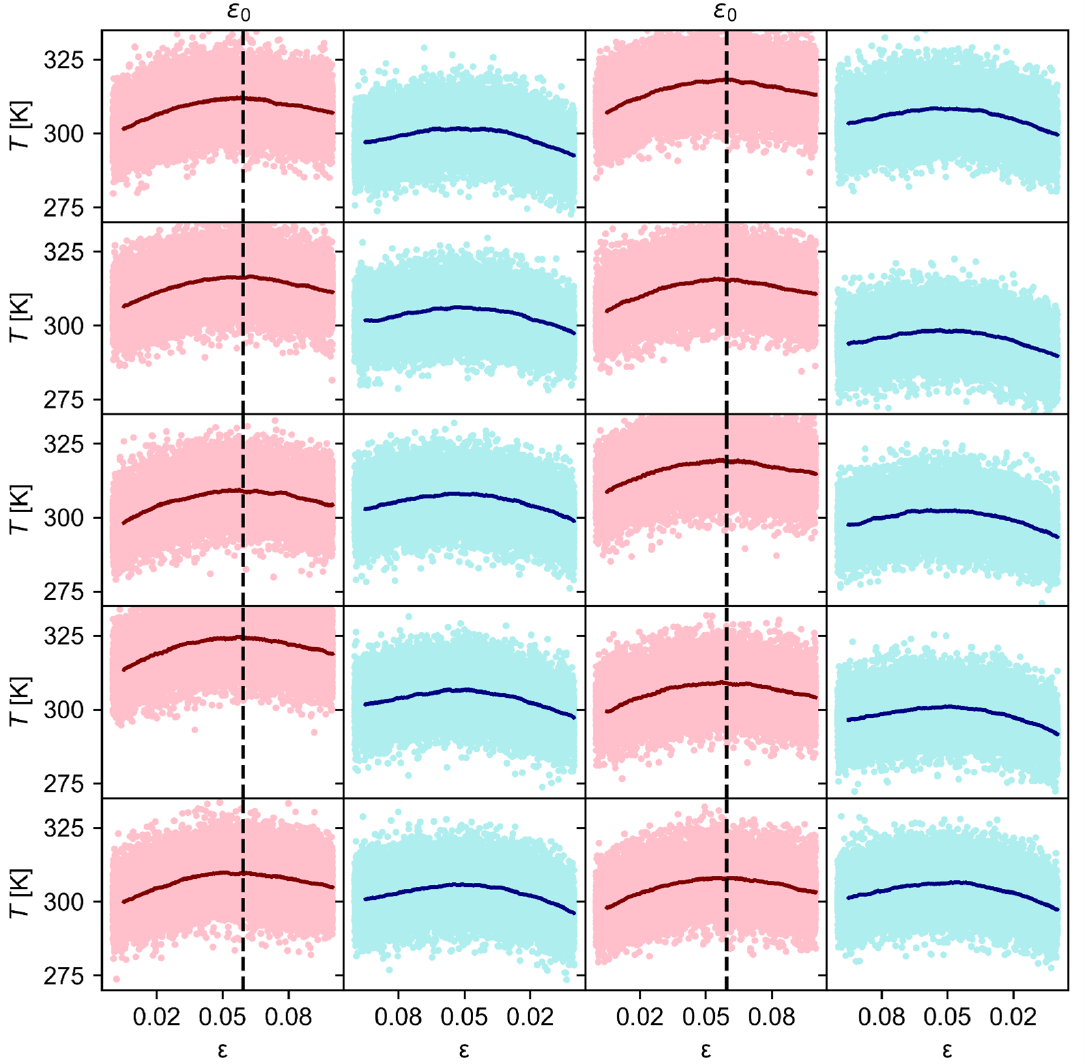}
    \caption{Ten pairs of temperature change profiles of expansion (red) and contraction (blue) paths of the thermodynamic cycles simulated with a zigzag $\gamma$-graphyne nanoribbon on a substrate. The strain $\varepsilon_0$ corresponds to the maximum temperature change during expansion.}
    \label{figS3.3}
\end{figure}

\renewcommand{\thefigure}{S3.4}
\begin{figure}
    \centering
    \includegraphics[width=0.8\linewidth]{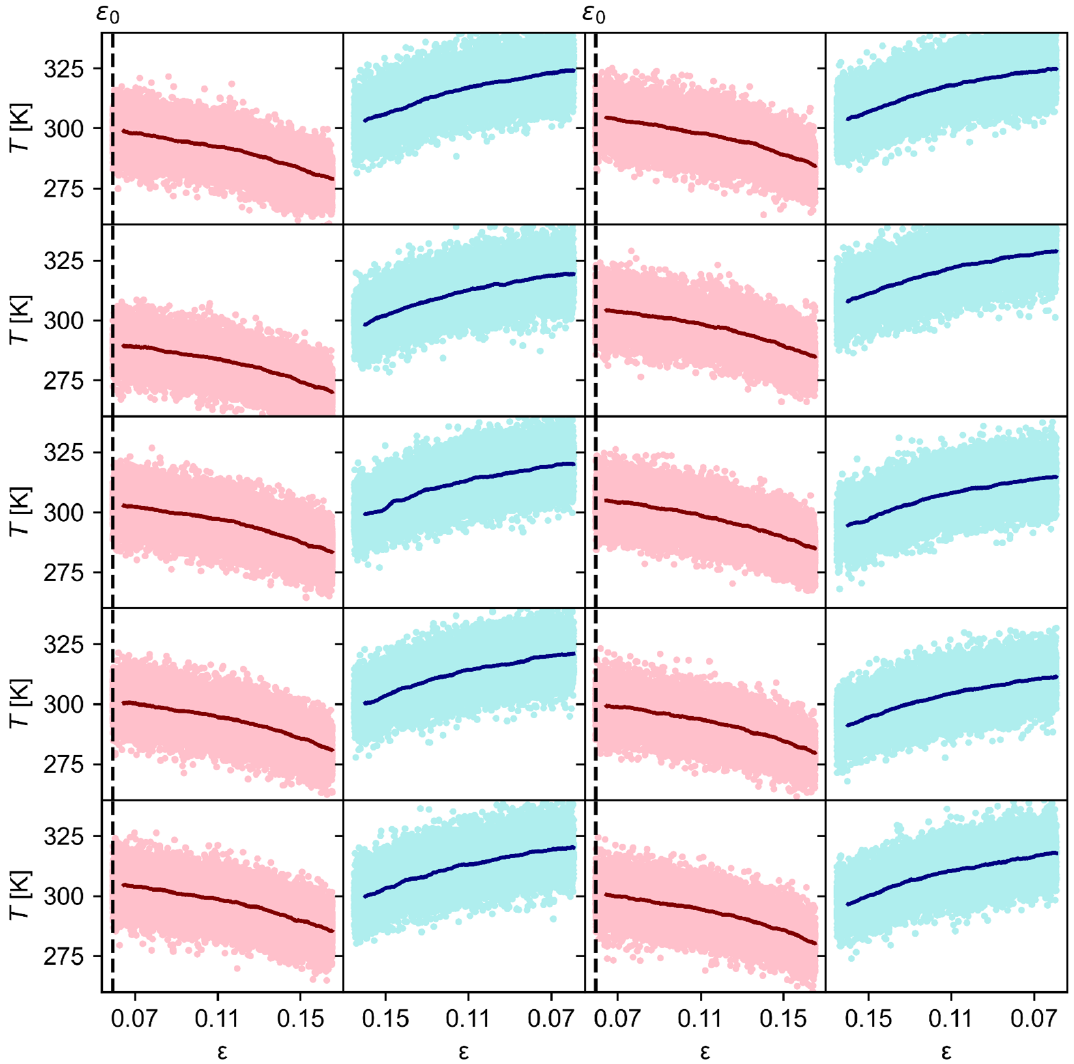}
    \caption{Ten pairs of temperature change profiles of expansion (red) and contraction (blue) paths of the thermodynamic cycles simulated with a zigzag $\gamma$-graphyne nanoribbon on a substrate with a pre-elongation of $\varepsilon_0$.}
    \label{figS3.4}
\end{figure}

\newpage

\section{S4. Molecular Dynamics Specific Heat Calculations}

Classical heat capacities of the $\gamma$-graphyne structures with periodic boundary conditions (PBC) and finite length were determined through a series of molecular dynamics (MD) simulations at fixed temperatures ranging from $T = 50$ K to $T = 700$ K. Given that the thermal expansion of graphyne is on the order of $10^{-5}$ K$^{-1}$ \cite{Hernandez2017DRM}, these simulations were carried out with fixed lengths. Figure \ref{figS4.1} illustrates the temperature and energy profiles for armchair and zigzag finite length and suspended structures. 

\renewcommand{\thefigure}{S4.1}
\begin{figure}
    \centering
    \includegraphics[width=\linewidth]{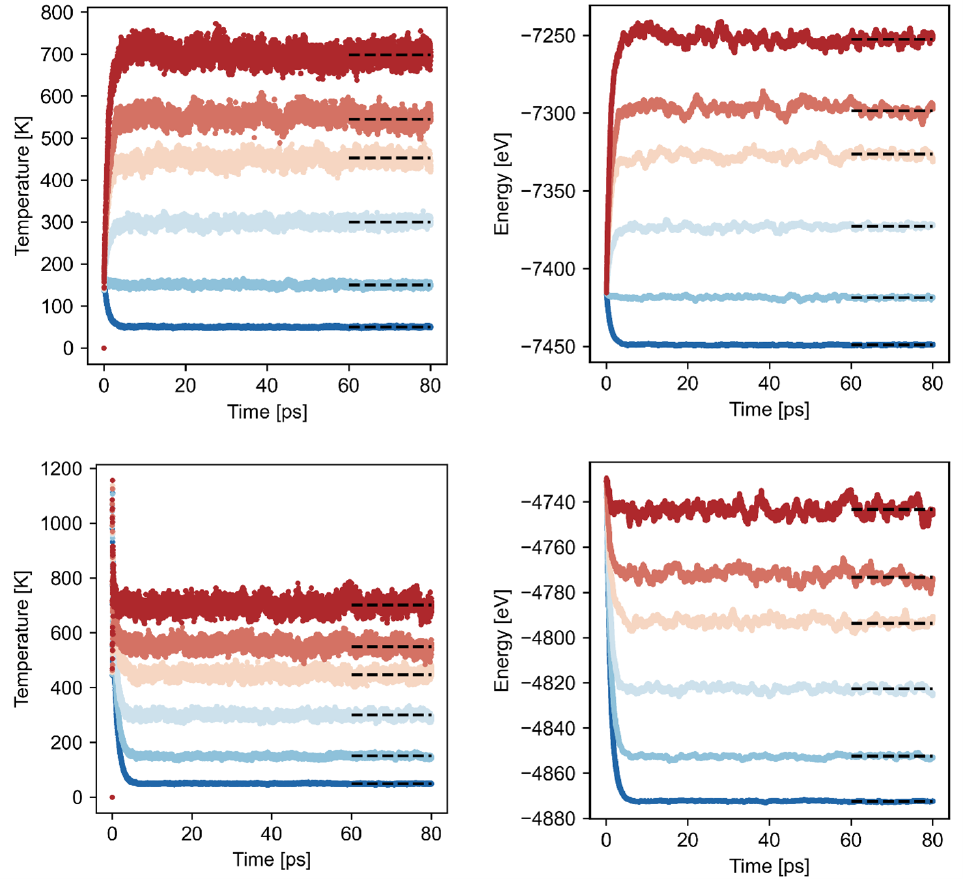}
    \caption{Temperature (left) and energy (right) profiles of the series of molecular dynamics simulations at different temperature values for armchair (top row) and zigzag (bottom row) finite length and suspended structures. The dashed lines indicate the equilibrium values taken to obtain the heat capacities of the structures further.}
    \label{figS4.1}
\end{figure}

Once the equilibrium values of energy versus temperature are obtained, the heat capacity can be directly estimated. Figure \ref{figS4.2} shows the results for the two examples shown in Figure \ref{figS4.1}. The obtained heat capacities are 0.198 and 0.304 eV/K. As we have the mass of each structure simulated in the present study, the specific heat of all structures can be calculated from the above heat capacities and are shown in Table \ref{tabS4.1}. 

\renewcommand{\thefigure}{S4.2}
\begin{figure}
    \centering
    \includegraphics[width=\linewidth]{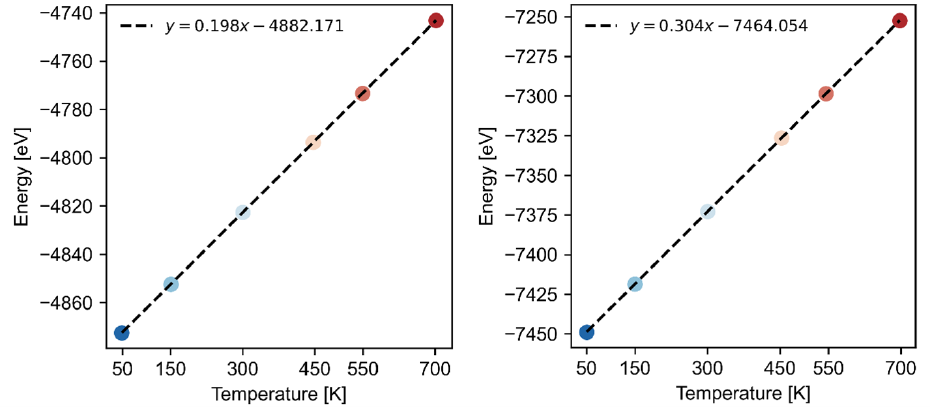}
    \caption{Equilibrium energy versus temperature for zigzag (left) and armchair (right) finite length and suspended structures. Dashed lines represent the linear fitting of the data.}
    \label{figS4.2}
\end{figure}

\renewcommand{\thetable}{S4.1}
\begin{table}
  \caption{Heat capacities, total mass of the free part of our structures (the part that is allowed to thermally fluctuates), the corresponding specific heats and the MD to real specific heat ratio, where the value of the real specific heat was obtained from density functional theory (DFT) calculations and is given by 0.4383 J/gK. The ratio will be used in equation (1) of the main text to obtain the real eC temperature change. }
  \label{tabS4.1}
    \begin{tabular}{P{2cm} l P{2cm}}\hline
        \multicolumn{3}{c}{\textbf{Heat Capacity [eV/K]}} \\\hline
         AC      &  &  ZZ             \\ \cline{1-1}\cline{3-3}
        0.304    &  & 0.198          \\\hline
    \end{tabular}
    \vspace{0.5cm}
 
    \begin{tabular}{ccccccccc}
        \hline\multicolumn{9}{c}{\textbf{Specific Heat {[}J/g $\cdot$ K{]}}}       \\\hline
          & \multicolumn{3}{c}{AC} &   & \multicolumn{3}{c}{ZZ} &   \\ \cline{2-4} \cline{6-8}
          & PBC     &    & FL      &   & PBC     &    & FL      &   \\ \cline{2-2} \cline{4-4} \cline{6-6} \cline{8-8}
          & 2.233   &    & 2.305   &   & 3.442   &    & 3.628   &   \\\hline
    \end{tabular}
    \vspace{0.5cm}
 
    \begin{tabular}{ccccccccc}
        \hline\multicolumn{9}{c}{\textbf{Total Mass [$\mathbf{\times 10^{-20}}$ g]}} \\\hline
          & \multicolumn{3}{c}{AC} &   & \multicolumn{3}{c}{ZZ} &   \\ \cline{2-4} \cline{6-8}
          & PBC     &    & FL      &   & PBC     &    & FL      &   \\ \cline{2-2} \cline{4-4} \cline{6-6} \cline{8-8}
          & 2.1807   &    & 2.1125   &   & 1.4148   &    & 1.3425   &   \\\hline
    \end{tabular}
    \vspace{0.5cm}
 
    \begin{tabular}{ccccccccc}
        \hline\multicolumn{9}{c}{\textbf{$\mathbf{c_{MD}/c_{REAL}}$}} \\\hline
          & \multicolumn{3}{c}{AC} &   & \multicolumn{3}{c}{ZZ} &   \\ \cline{2-4} \cline{6-8}
          & PBC     &    & FL      &   & PBC     &    & FL      &   \\ \cline{2-2} \cline{4-4} \cline{6-6} \cline{8-8}
          & 5.09468   &    & 5.25896   &   & 7.85307   &    & 8.27744   &   \\\hline
    \end{tabular}

\end{table}

\end{document}